\documentclass[fleqn,usenatbib]{mnras}
\usepackage{newtxtext,newtxmath}
\usepackage[T1]{fontenc}
\usepackage{xcolor}

\DeclareRobustCommand{\VAN}[3]{#2}
\let\VANthebibliography\thebibliography
\def\thebibliography{\DeclareRobustCommand{\VAN}[3]{##3}\VANthebibliography}

\usepackage{graphicx}
\usepackage{amsmath}
\usepackage{float}
\usepackage{CJK}
\usepackage{siunitx}

\newcommand\fred{SCR~J1845--6357}
\newcommand\tyc{TYC~9078--815--1}
\newcommand\gaiasource{Gaia~6438561254116213632}
\newcommand\cdstar{CD--64~1208}
\newcommand\ftoct{V*~FT~Oct}
\newcommand\askapj{ASKAP~J212525.4-813829}
\newcommand\cfoct{V*~CF~Oct}
\newcommand\pulsar{PSR~J1933--6210}

\newcommand\fstar{UCAC~048-020372}
\newcommand\ucac{UCAC~053-023945}

\newcommand{\changes}[1]{\textcolor{black}{#1}}

\title[Radio transients and variables in DWF O10]{Radio Transients and Variables in the Tenth Deeper, Wider, Faster Observing Run}

\author[D. Dobie et al.]{D. Dobie,$^{1,2}$\thanks{E-mail: ddobie@swin.edu.au}
J. Pritchard,$^{3,2,4}$
Y. Wang (王远明),$^{3,2,4}$ 
L. W. Graham,$^{1}$ 
J. Freeburn,$^{1,2}$ 
H. Qiu (邱昊),$^{5}$ 
\newauthor
T. R. White,$^{3}$ 
A. O'Brien,$^{6}$
E. Lenc,$^{4}$ 
J. K. Leung,$^{3,2,4}$ 
C. Lynch,$^{7,8}$ 
Tara Murphy,$^{3,2}$ 
A. J. Stewart,$^{3}$ 
\newauthor
Z. Wang (王子腾),$^{3,2,4}$ 
A. Zic,$^{9,4}$ 
T. M. C. Abbott,$^{10}$ 
C. Cai (蔡策),$^{11,12}$ 
J. Cooke,$^{1,2}$ 
M. Dobiecki,$^{1}$ 
\newauthor
S. Goode,$^{1,2}$ 
S. Jia (贾淑梅)$^{11}$, 
C. Li (李承奎)$^{11}$, 
A. M\"oller,$^{1,2}$, 
S. Webb,$^{1,2}$ 
J. Zhang (张洁莱),$^{1,2}$ 
\newauthor
S. N. Zhang(张双南)$^{11,12}$ 
\\
$^{1}$ Centre for Astrophysics and Supercomputing, Swinburne University of Technology, Hawthorn, Victoria, Australia\\
$^{2}$ ARC Centre of Excellence for Gravitational Wave Discovery (OzGrav), Hawthorn, Victoria, Australia\\
$^{3}$ Sydney Institute for Astronomy, School of Physics, University of Sydney, NSW 2006, Australia\\
$^{4}$ Australia Telescope National Facility, CSIRO, Space and Astronomy, PO Box 76, Epping, NSW 1710, Australia\\
$^{5}$ SKA Observatory, Jodrell Bank, Lower Withington,
Macclesfield, SK11 9FT, UK\\
$^{6}$ Department of Physics, University of Wisconsin-Milwaukee, P.O. Box 413, Milwaukee, WI 53201, USA\\
$^{7}$ International Centre for Radio Astronomy Research, Curtin University, Perth, WA 6845, Australia\\
$^{8}$ ARC Centre of Excellence for All Sky Astrophysics in 3 Dimensions (ASTRO-3D)\\
$^{9}$ School of Mathematical and Physical Sciences, and Research Centre in Astronomy, Astrophysics and Astrophotonics, Macquarie University, NSW 2109, Australia\\
$^{10}$ Cerro Tololo Inter-American Observatory, NSF's National Optical-Infrared Astronomy Research Laboratory, Casilla 603, La Serena,
Chile\\
$^{11}$ Key Laboratory of Particle Astrophysics, Institute of High Energy Physics, Chinese Academy of Sciences, 19B Yuquan Road, Beijing 100049, China\\
$^{12}$ University of Chinese Academy of Sciences, Chinese Academy of Sciences, Beijing 100049, China
}

\date{Accepted XXX. Received YYY; in original form ZZZ}
\pubyear{2022}

\begin{document}
\begin{CJK*}{UTF8}{gbsn}
\label{firstpage}
\pagerange{\pageref{firstpage}--\pageref{lastpage}}
\maketitle

\begin{abstract}
\changes{The Deeper, Wider, Faster (DWF) program coordinates observations with telescopes across the electromagnetic spectrum, searching for transients on timescales of milliseconds to days.} The tenth \changes{DWF} observing run was carried out \changes{in near real-time} during September 2021 and consisted of six consecutive days of observations of \changes{the NGC\,6744 galaxy group and a field containing the repeating fast radio burst}  FRB\,190711 with the Australian Square Kilometre Array Pathfinder, the Dark Energy Camera, the Hard X-ray Modulation Telescope and the Parkes 64\changes{m ``Murriyang''} radio telescope. In this work we present the results of an image-domain search for transient, variable and circularly polarised sources carried out with ASKAP using data from the observing run, along with test observations prior to the run and follow-up observations carried out during and after the run. We identified eight variable radio sources, consisting of one pulsar, six stellar systems (five of which exhibit circularly polarised emission) and one previously uncatalogued source. Of particular interest is the detection of pulses from the ultra-cool dwarf \fred\ with a period of $14.2\pm 0.3$\,h, in good agreement with the known optical rotation period, making this the slowest rotating radio-loud ultra-cool dwarf discovered.
\end{abstract}

\begin{keywords}
radio continuum: stars -- radio continuum: transients -- stars: low-mass -- stars: flare -- polarization
\end{keywords}

\end{CJK*}
\section{Introduction}
\label{sec:intro}
The Deeper, Wider, Faster program \citep[DWF\changes{;}][]{2019IAUS..339..135A} coordinates simultaneous observations between telescopes across the entire electromagnetic spectrum to study astrophysical transients on timescales of milliseconds to days.
The tenth DWF observing run took place from 2021 September 5--10 and consisted of simultaneous observations with the Australian Square Kilometre Array Pathfinder \citep[ASKAP;][]{2021PASA...38....9H}, the Parkes 64m \changes{``Murriyang''} radio-telescope, the Dark Energy Camera (DECam) and the Hard X-ray Modulation Telescope (HXMT) carried out on a daily cadence. We observed \changes{fields containing the NGC\,6744 galaxy group and the repeating fast radio burst FRB\,190711, for approximately 1.5 and 3\,h per night respectively.}

ASKAP has been used in previous DWF runs (Cooke et al. in prep.), but only to search for Fast Radio Bursts (FRBs) \changes{in high time resolution data. However, ASKAP has previously been used for similar multi-wavelength studies, for example a targeted campaign of Proxima Centauri \citep[][]{2020ApJ...905...23Z} and co-observing with the Transiting Exoplanet Survey Satellite \citep[][]{2022MNRAS.tmp.2052R}. The tenth DWF observing run is the first time that ASKAP interferometric visibilities have been recorded as part of DWF observing, enabling image-domain searches for transients on timescales of minutes to days.}

Past widefield searches for transient and variable sources at gigahertz frequencies have generally observed on timescales of weeks to years \citep[e.g.][]{2013ApJ...768..165M,2016ApJ...818..105M,2018ApJ...866L..22L,2020ApJ...903..116A,2022MNRAS.510.3794D}, while those using faster cadences tend to consist of narrow-field observations targeting known sources \citep[e.g.][]{2011MNRAS.415....2B,2011ApJ...728L..14B,2022MNRAS.512.5037D}. The $30\,\deg^2$ field-of-view provided by ASKAP has allowed us to perform a widefield search on a high cadence and hence, the observations presented in this work provide a window into a previously unexplored part of the radio transient parameter space.

The cadence of these observations means that our search was not sensitive to the majority of extragalactic radio transients, which only exhibit intrinsic variability on day--week timescales during the very early-time rise in their light curves. Some may produce shorter-lived emission \citep[e.g. reverse shocks from gamma-ray bursts][]{2013ApJ...776..119L,2018Galax...6..103L}, but those compose a small subset of already rare events. \changes{Instead, our observations are most sensitive to variability on timescales of minutes to days, such as the refractive scintillation of extragalactic sources. This variability is characterised by quasi-random oscillations on timescales of days with an amplitude of tens of percent \citep[][]{1998MNRAS.294..307W}. The properties of this variability can be used to study both the objects producing the emission \citep[e.g.][]{2022MNRAS.512.5358R} and the properties of the intervening material \citep[e.g.][]{2021MNRAS.502.3294W}.}

\changes{Variants of neutron star and systems containing them, such as pulsars, magnetars and X-ray binaries (XRBs), are known to produce variable radio emission on timescales of seconds to days. The emission from all of these systems is generally coherent, and can also be highly circularly polarised. Their variability can be extrinsic (e.g. the scintillation of any compact source), or intrinsic, such as pulsar mode changes \citep{2016MNRAS.456.3948H}, magnetar outbursts \citep{2011ASSP...21..247R} and XRB flares \citep[][]{2009MNRAS.396.1370F}. The origin of Galactic Centre Radio Transients \citep{2002AJ....123.1497H,2005Natur.434...50H,2009ApJ...696..280H} is as yet unknown, but may also be related to neutron stars. Recent work has also discovered a number of similar radio variables in or near the Galactic plane that are comparable to the aforementioned classes, but do not their standard paradigm \citep{2021ApJ...920...45W,2022Natur.601..526H,2022MNRAS.516.5972W}. However, the primary goal of the DWF program is the multi-wavelength detection of an extragalactic FRB, and the fields observed in this work were chosen to minimise both the potential Galactic contribution to the FRB dispersion measure, and extinction at optical wavelengths. Hence, we expect the detection rates of these neutron star related objects, which are predominantly situated in the Galactic plane, to be low.}

\changes{Some stars can produce bursts or flares \citep[see][for an overview]{2002ARA&A..40..217G} that are often coherent, last seconds--minutes, and can also be highly circularly polarised \citep[][]{1985ARA&A..23..169D}. Variable quiescent emission has also been observed from stellar systems, has been hypothesised to originate from the combined contribution of multiple small flares \citep{2011ASPC..448..455F}, but has also been demonstrated to be caused by rotational modulation in some instances \citep[e.g.][]{2006A&A...458..831L}. While the underlying distribution of stars is Galactic, the distribution of those detectable in this survey is sensitivity limited and we therefore we don't expect the high Galactic latitude of these observations to impact our ability to detect them.}

In this work, we present a search for transient and variable sources in the ASKAP observations carried out during the tenth DWF run. This work is focused on ``slow'' (timescales longer than minutes) radio variability and hence we do not discuss the search for FRBs with ASKAP or Parkes, nor the optical transient search carried out with DECam. We have carried out two radio variability studies \changes{-- the first using each day's observations; and the second on 15-minute images created from each observation. We have also searched for circularly polarised sources, as a substantial fraction of the variable radio emission these observations are sensitive to is expected to be circularly polarised.}

\begin{table}
    \centering
    \begin{tabular}{cccrrc}
    \hline\hline
    Field & SBID & Start Time & Duration & $\sigma_{\rm RMS}$\\
     & & (UTC) & (hh:mm:ss) & ($\mu$Jy)\\
    \hline
    FRB190711 & 31377 & 2021-08-28 15:45:15 & 03:22:50 & 72\\
    FRB190711 & 31495 & 2021-08-31 07:39:48 & 03:21:01 & 119\\
    FRB190711 & 31585 & 2021-09-02 07:03:48 & 03:21:21 & 65\\
    FRB190711 & 31652 & 2021-09-03 10:48:04 & 03:23:22 & 59\\
    FRB190711 & 31701 & 2021-09-04 06:25:30 & 00:48:24 & 131\\
    FRB190711 & 31702 & 2021-09-04 07:19:58 & 03:21:12 & 64\\
    FRB190711 & 31751 & 2021-09-05 06:22:03 & 03:21:28 & 65\\
    FRB190711 & 31822 & 2021-09-06 06:21:53 & 03:21:20 & 65\\
    FRB190711 & 31883 & 2021-09-07 06:25:26 & 03:21:21 & 67\\
    FRB190711 & 31945 & 2021-09-08 06:22:00 & 03:21:27 & 72\\
    FRB190711 & 32016 & 2021-09-09 06:26:48 & 03:23:24 & 67\\
    FRB190711 & 32036 & 2021-09-10 06:22:01 & 03:23:37 & 66\\
    \rule{0pt}{3ex}  
    NGC6744 & 31349 & 2021-08-27 16:37:24 & 01:21:05 & 87\\
    NGC6744 & 31584 & 2021-09-02 05:41:31 & 01:22:15 & 107\\
    NGC6744 & 31640 & 2021-09-03 05:15:00 & 01:22:28 & 107\\
    NGC6744 & 31651 & 2021-09-03 09:22:47 & 01:25:15 & 92\\
    NGC6744 & 31700 & 2021-09-04 05:21:59 & 01:03:30 & 125\\
    NGC6744 & 31750 & 2021-09-05 04:59:58 & 01:22:00 & 110\\
    NGC6744 & 31821 & 2021-09-06 04:59:57 & 01:21:53 & 108\\
    NGC6744 & 31882 & 2021-09-07 05:02:47 & 01:22:37 & 112\\
    NGC6744 & 31944 & 2021-09-08 05:00:03 & 01:21:53 & 119\\
    NGC6744 & 32015 & 2021-09-09 05:00:03 & 01:26:43 & 114\\
    NGC6744 & 32018 & 2021-09-09 12:09:26 & 07:01:07 & 45\\
    NGC6744 & 32035 & 2021-09-10 05:00:01 & 01:21:58 & 115\\
    NGC6744 & 32039 & 2021-09-10 12:08:40 & 07:01:19 & 42\\
    NGC6744 & 32235 & 2021-09-19 05:20:01 & 10:01:01 & 38\\
    \hline\hline
    \end{tabular}
    \caption{Details of the ASKAP observations carried out as part of the tenth DWF observing run, including the median image noise, $\sigma_{\rm RMS}$. We also provide the scheduling block (SBID) which can be used to access the data via the ASKAP Science Data Archive.}
    \label{tab:obs_descrip}
\end{table}

\section{Searching for Transients and Variables with ASKAP}
\subsection{Observations} All ASKAP observations were carried out with the \textsc{closepack36} beam footprint \citep[see Fig. 20 of ][]{2021PASA...38....9H} centered on 19:08:00.000,$-$64:30:00.00 and 22:00:20.447,$-$79:55:21.84 for the NGC\,6744 and FRB\,190711 fields\changes{,} respectively\changes{,} and a central frequency of 943\,MHz with 288\,MHz of bandwidth.

We carried out nightly observations of both fields from 2021 September 5 to 10, targeting the NGC\,6744 field for approximately 80 minutes per night, and the FRB\,190711 field for 3 hours and 20 minutes per night. ASKAP was on-field for the entirety of the run, but also carried out a number of \changes{observations} of both fields in the week prior to the run \changes{to test both the operation and scheduling of the telescope}. Additionally, we carried out three target-of-opportunity observations of the NGC\,6744 field in response to our discovery of radio emission from \fred\ (see Section \ref{subsec:speedy_fred}), which we also include in this analysis. A \changes{summary} of the observations reported in this paper can be found in Table \ref{tab:obs_descrip}.

Each observation was calibrated and imaged using the default ASKAPsoft pipeline \citep{2017ASPC..512..431W}, which produces calibrated visibilities, full-polarisation images \changes{and} noise maps, \changes{as well as} Stokes I source catalogues for each observation.

\begin{table*}
    \centering
    \begin{tabular}{lccccccccc}
        \hline\hline
        Name & Right Ascension & Declination & $\eta$ & $V$ & Polarisation & Fast Imaging & Gaia parallax & Optical period\\
         & (hh:mm:ss) & (dd:mm:ss) &  &  &  &  & (mas) & (days)\\
        \hline
        \fred &    18:45:13.73 & -63:57:34.9 &  96 &  1.0 & Yes & Yes & $249.7\pm 0.1$ & $0.588\pm 0.008$ \\
        \ucac &    21:07:28.58 & -79:26:27.3 &   3.4 &  0.55 & Yes & No & $14.44\pm0.02$ & $17.16\pm0.2$\\
        \gaiasource &    19:04:03.68 & -64:00:12.1 &   5.9 & 30 & Yes & Yes & $2.27\pm 0.05$ & --\\
        \fstar &    22:01:25.30 & -80:34:35.6 &   4.27 &  1.3 & Yes & No & $13.32\pm 0.03$ & $9.5\pm 1.4$\\
        \cfoct &    20:49:37.54 & -80:08:00.6 &  79 &  0.42 & No & No & $4.45\pm 0.06$ & 19.90--20.45\\
        \askapj &    21:25:25.35 & -81:38:28.7 &   3.5 &  0.17 & No & Yes & -- & --\\
        \pulsar &    19:33:32.39 & -62:11:46.9&  83 &  0.64 & No & Yes & -- & --\\
        \cdstar &    18:45:37.12 & -64:51:49.3 & 170 &  0.40 & Yes & Yes & $35.2\pm0.2$ & $0.354\pm0.004$\\
        \hline\hline
    \end{tabular}
    \caption{Properties of the eight variable sources discovered in this work. $\eta$ and $V$ are the variability metrics calculated by the VAST pipeline (see Section \ref{subsec:pipeline-search} for details). ``Polarisation'' denotes sources that were found by the circular polarisation search (Section \ref{subsec:pol-search}) and ``Fast Imaging'' denotes sources that were found by the fast-imaging search (Section \ref{subsec:fast-search}). Optical periods are measured using TESS observations (Section \ref{subsec:tess}) except for \cdstar\ where we report the measurement from \citet{2011A&A...533A..30G} and \cfoct\ where we report the period range measured by \citet{2012MNRAS.420.2539B}.}
    \label{tab:variables}
\end{table*}

\subsection{VAST pipeline}
\label{subsec:pipeline-search}
We carried out a search for variable sources with the Variables And Slow Transients (VAST) pipeline\footnote{\url{https://vast-survey.org/vast-pipeline/}} \citep{2021PASA...38...54M,2022ASPC..532..333P}, using the Stokes I images, noise maps and source catalogues produced by ASKAPsoft. As in \citet{2022MNRAS.510.3794D} we performed source association with a de Ruiter radius of 5.68 \citep[][]{2011PhDT.......427S} and used the equations of \citet{1997PASP..109..166C} to recalculate the source fitting uncertainty estimates.

We applied the following cuts to minimise the number of image artefacts contaminating our sample:
\begin{itemize}
    \item \changes{a} ratio of integrated to peak flux density $<1.5$;
    \item \changes{n}o relations, i.e. it is not associated with any other source\footnote{See \url{https://vast-survey.org/vast-pipeline/design/association/##relations}};
    \item \changes{d}istance to nearest source $>1\,$\arcmin;
    \item \changes{s}ignal-to-noise ratio $>5$.
\end{itemize}

We note that while these cuts potentially exclude real variable sources, in the absence of dedicated artefact detection algorithms (which are currently under development) they are necessary in order to remove the substantial number of artefacts that are associated with bright sources in our data. 

For sources detected in at least two observations, we carried out a standard $\eta$--$V$ variability search \changes{where}
\begin{equation}
    V = \frac{1}{\overline{S}} \sqrt{\frac{N}{N-1} \left(\overline{S^2}-\overline{S}^2\right)},
\end{equation}
\changes{is the flux density variability relative to the weighted mean and}
\begin{equation}
    \eta = \frac{N}{N-1}\left(\overline{wS^2}-\frac{\overline{wS}^2}{\overline{w}}\right),
\end{equation}
\changes{quantifies the statistical significance of that variability. In both the above equations} $N$ is the number of datapoints, $S$ is the flux density\changes{, $w$ is the inverse uncertainty and overbars denote the mean of a quantity}. These statistics were originally defined by \citet{2015A&C....11...25S} and a full description of the search technique we used can be found in \citet{2022MNRAS.510.3794D}. We then manually inspected all candidates, removing all of those that were likely imaging artefacts or were quasi-random, low-amplitude \changes{($V \lesssim 0.3$)} oscillations around a mean flux density (as expected from refractive scintillation). All sources with a single detection were also manually inspected. \changes{From 2021-09-07 onwards this search was carried out the morning after each observation, with a list of candidates compiled before the next day's observations began. This is the first time an image domain radio transient search has been carried out in near-real-time, and ensured that we were able to trigger multi-wavelength follow-up observations of any candidates as necessary.}

Eight sources passed our vetting process\footnote{\pulsar\ does not pass the criteria in the final pipeline run due to a spurious single-epoch noise spike offset from the source by 0.43\arcmin. \gaiasource\ does not pass the criteria in the final run due to a faint source offset by 0.43\arcmin\ detected in the 3 deep epochs. However, both \changes{passed our selection criteria} in earlier pipeline runs and hence we include them in this work.}. Crossmatching with SIMBAD, NED and the TESS Input Catalogue, we find that six are coincident with known stellar systems, one is coincident with a known pulsar and the remaining variable has no known multi-wavelength counterparts. These sources and their variability metrics are listed in Table \ref{tab:variables} and discussed in detail in Section \ref{subsec:sources-of-interest}.

\subsection{Polarisation Search}
\label{subsec:pol-search}
We performed a circular polarisation search to identify sources with a high fractional circular polarisation $f_p = |S_V|/S_I$ using a similar technique to that presented by \citet{2021MNRAS.502.5438P}. \changes{This search does not directly probe variability but, as we note in Section \ref{sec:intro}, the variable sources that these observations are predominantly sensitive to can often produce circularly polarised emission. Hence, this search serves as a way to detect variable sources of interest that may have been missed by our other searches (e.g. highly polarised pulses that are detected at low significance).}

We first extracted Gaussian source components from the Stokes V images using the {\sc selavy} \citep{2012PASA...29..371W} source finder package, using the same, standard {\sc selavy} settings as used in ASKAPsoft for the Stokes I images. We ran source extraction twice, with the second pass running over the inverted images, to extract sources with both positive and negative Stokes V emission. We then generated associations between the Stokes I and V components by performing a many-to-many crossmatch within a search radius of \SI{6}{\arcsec}. Many-to-many association allows for all permutations of I-V matches between multi-component sources, ensuring no candidates are missed due to incorrect association with near neighbours or selavy artefacts. 

We then crossmatched the resulting list of I-V associations across each epoch, and required that candidates contained at least one epoch with $f_p > 0.06$ in order to remove I-V associations attributed to polarisation leakage \citep[which can be as high as $\sim 2$~per~cent;][]{2021MNRAS.502.5438P}. This resulted in a final set of 51 candidates in the FRB190711 field and 74 candidates in the NGC6744 field. We manually inspected each candidate to identify associations to imaging artefacts caused by bright or extended sources and associations to spurious noise peaks in Stokes V, and identified five sources coincident with stars (see Table \ref{tab:variables}.

\subsection{Fast-imaging Search}
\label{subsec:fast-search}
We carried out a search for rapid variable sources on 15-minute timescales with the VAST fast-imaging pipeline (Wang Y. et al. in prep) using the self-calibrated visibilities produced by ASKAPsoft.

We processed each of the 36 beams separately, and all imaging was carried out using \textsc{CASA}. For each beam, we made an independent sky model image using multi-scale multi-frequency synthesis with two Taylor terms from the self-calibrated visibilities \changes{from the full observation}. 
We performed a deep clean (10\,000 iterations) using Briggs weighting with robustness of 0.5 to provide a compromise between resolution and sensitivity, and achieved a mean residual RMS noise of $\sim$70\,$\mu$Jy\,beam$^{-1}$ for \changes{the} FRB190711 fields\footnote{Excluding SB31701 which is too short to conduct a \changes{fast-imaging search because each source would have 3 measurements at most.}} and a mean residual RMS noise of $\sim$100\,$\mu$Jy\,beam$^{-1}$ for \changes{the} NGC6744 fields.
We then converted each sky model image to model visibilities, and subtracted the model visibilties from the calibrated visibilities. 
These model-subtracted visibilities were imaged at 15-min timesteps using the same weighting parameters, generating a series of short 15-min images for variability analysis.

For each beam, we generated a source catalog from the independent sky model restored image using \textsc{Aegean} \citep{2012MNRAS.422.1812H,2018PASA...35...11H} using a $6\sigma$ detection threshold. For each source in the sky model, we converted the global coordinate to the pixel position in each image, and at each position measured
\begin{itemize}
    \item the deep flux density $S_\mathrm{d}$ on the sky model restored image; 
    \item the residual flux density $S_\mathrm{r}$ on the sky model residual image; and
    \item the peak flux density $S_{i, \mathrm{s}}$ on the $i$-th short 15-min image. 
\end{itemize}
The $i$-th data point of the intra-observation light curve is then given by
\begin{equation}
    S_i = S_{i, \mathrm{s}} + S_\mathrm{d} - S_\mathrm{r}
\end{equation}

We then carried out a variability analysis, using the modulation index $m$ to characterize the magnitude of variability, and reduced chi-squared $\chi^2$ to measure its statistical significance \citep[See Section~3.2 of][]{2021MNRAS.502.3294W}. We selected candidates with a reduced chi-squared $\chi^2>2.0\sigma_{\chi^2}$ (where $\sigma_{\chi^2}$ is the standard deviation measured by fitting a Gaussian function to the distributions of reduced chi-squared in logarithmic space) and a modulation index $m>5\%$ \citep[for ruling out potential bright sources of which high chi-square value are dominated by high self-noises; see][]{1989AJ.....98.1112K}. 
We also ruled out potential extended sources with a ratio of integrated to peak flux density $>1.5$, and sources more than one full-width half maximum (\changes{FWHM;} $\sim 0.8\,\deg$) from the beam centre.
We manually inspected each candidate, including comparing its light curve with the light curve of the same source as observed by neighbouring beams to rule out the variability being caused by instrumental effects. This search ultimately revealed five variable sources as described in Table~\ref{tab:variables}.

\begin{table}
    \centering
    \begin{tabular}{rcc}
        \hline\hline
        Energy range (keV) & \multicolumn{2}{c}{Upper limit  (erg\,cm$^{-2}$\,s$^{-1}$)}\\
         & FRB190711 & NGC 6744 \\
        \hline
        1--10 & $8.9\times 10^{-10}$ & $5.6\times 10^{-10}$\\
        10--30 & $1.0\times 10^{-9}$ & $9.9\times 10^{-10}$\\
        28-250 & $1.4\times 10^{-9}$ & $1.5\times 10^{-9}$\\
        \hline\hline
    \end{tabular}
    \caption{Constraints (calculated as $3\sigma$ upper limits) on transient X-ray emission in the FRB190711 and NGC 6744 fields during our observations. The sensitivity of each detector is time variable and these constraints correspond to a typical period during our observations.}
    \label{tab:hxmt_limits}
\end{table}

\section{Multi-wavelength Observations}
\subsection{Simultaneous}
\subsubsection{Dark Energy Camera}
\label{subsec:decam}
We carried out simultaneous observations of both fields with the Dark Energy Camera \citep[DECam;][]{2015AJ....150..150F}, with a 3\,$\deg^2$ footprint ($\sim 2.5\deg^2$ effective field-of-view accounting for chip gaps) centered on 19:07:58.52 $-$64:30:00.4 and 21:57:20.87 $-$80:17:00.8 for the NGC\,6744 and FRB\,190711 fields respectively. We used the standard DWF observing strategy of 20\,s $g$-band exposures with a typical readout of 30\,s. We achieved a typical depth of 22.8\,mag and a typical seeing of 1.6\arcsec. An overview of the typical data reduction process can be found in \citet{2021MNRAS.506.2089W} and \citet{2020MNRAS.491.5852A}. An optical transient search was carried out in real-time, but discussion of it is beyond the scope of this work, and will instead be presented in a future manuscript.

\subsubsection{Hard X-ray Modulation Telescope}
\label{subsec:hxmt}
The Hard X-ray Modulation Telescope \citep[HXMT;][]{2020SCPMA..6349502Z} consists of three separate instruments -- the High Energy X-ray Telescope  \citep[HE; 27--250\,keV;][]{2020SCPMA..6349503L}, the Medium Energy X-ray Telescope  \citep[ME; 10--35\,keV;][]{2020SCPMA..6349504C} and the Low Energy X-ray Telescope \citep[LE; 1--10\,keV;][]{2020SCPMA..6349505C}. The instruments are co-aligned and were centered on 19:08:00 $-$64:30:00 and 21:57:40.8 $-$80:21:36 for the NGC\,6744 and FRB\,190711 fields respectively. The effective fields of view are $1.1\times5.7$\,deg$^2$ (HE), $1\times4$\,deg$^2$ (ME) and $1.6\times6$\,deg$^2$ (LE) although the sensitivity across the field is not uniform. Observations were carried out simultaneously with ASKAP, although we note that only \gaiasource\ and \fstar\ were within the HXMT field of view.

We do not detect any transient emission across either field in any instrument using the search technique described by \citet{2020A&A...642A.160G,2020A&A...637A..69G}. While the sensitivity of the telescope is not static, we place constraints on transient X-ray emission in the field by estimating a typical upper limit for each field and detector across the whole observing run. Table \ref{tab:hxmt_limits} shows these constraints as 3$\sigma$ upper limits.

\subsection{Archival}
\subsubsection{Gaia}
To determine precise distances for the sources we have identified in the three searches, we crossmatched the seven non-pulsar sources with the Gaia Data Release 3 catalogue \citep[DR3;][]{2016A&A...595A...1G,2022arXiv220800211G}. We searched for all catalogued sources within 10\arcsec of the radio source and then calculated the proper motion corrected positions at 2021-09-05. Six sources had Gaia counterparts with proper motion corrected positions in agreement with the radio position to within uncertainties. We list the parallax of each source in Table \ref{tab:variables}.

\subsubsection{Transiting Exoplanet Survey Satellite}
\label{subsec:tess}
We used data from the Transiting Exoplanet Survey Satellite (TESS) to search for optical periodicity from the six stellar sources. Two sources (\cfoct, \ucac) have light curves accessible via the MAST archive, while \cdstar\ is not within the footprint of any TESS observations. The remaining three sources with optical counterparts are within the TESS footprint, but do not have pre-existing light curves. \gaiasource\ is too close to \tyc\ to have reliable photometry, but for the other two sources (\fred\ and \fstar) we manually extract light curves from the TESS Full-Frame-Images using the standard {\sc lightkurve} method of threshold masking and background subtraction\footnote{\url{https://docs.lightkurve.org/tutorials/2-creating-light-curves/2-1-cutting-out-tpfs.html}} at the Gaia proper-motion corrected position at the time of observation.

Qualitatively, all five sources with TESS light curves exhibit clear optical periodicity. We generated a Lomb-Scargle periodogram of each light curve and report the period \changes{(Table \ref{tab:variables})} corresponding to the peak power and the associated uncertainty using the \changes{FWHM} of the peak. For sources that were observed in multiple TESS sectors we performed the aforementioned calculations on a sector-by-sector basis and then combined the resulting estimate of the period.

\begin{figure*}
    \centering
    \includegraphics{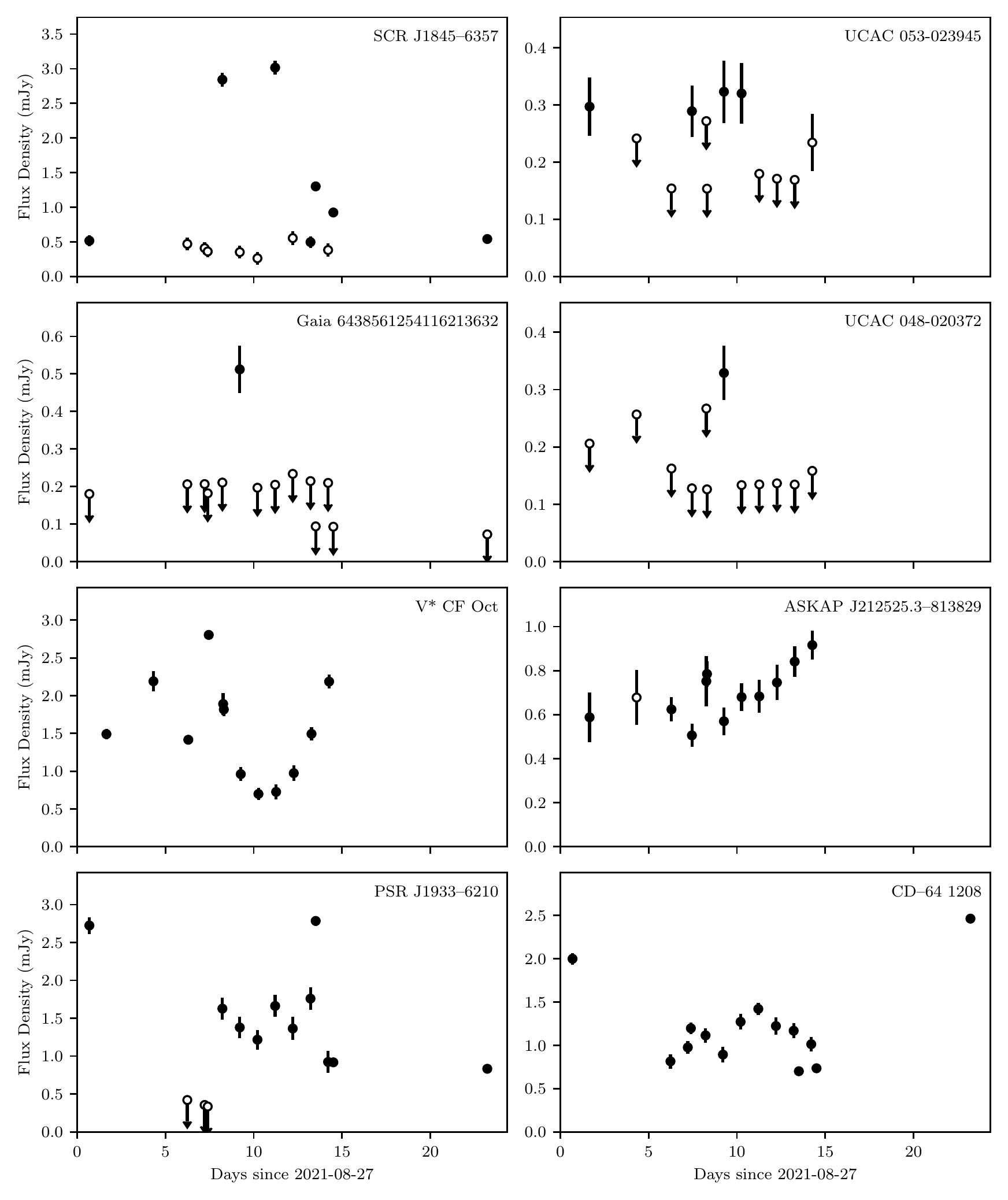}
    \caption{ASKAP light curves of the 8 variable sources found in this work. After manually inspecting the images of all 8 variables, we have determined that the non-detections of \fred\ and \askapj\ and one non-detection of \ucac053\ are sub-threshold detections and we therefore show the forced fit flux measurements as calculated by the VAST pipeline as measurements with open markers. Non-detections of other sources are shown as $3\sigma$ upper limits, also with open markers.}
    \label{fig:star_light_curves}
\end{figure*}

\section{Radio sources of interest}
\label{subsec:sources-of-interest}
We found eight variable radio sources in total, with all passing the VAST pipeline variability criteria \changes{and five passing the fast-imaging variability criteria. In addition, our polarisation search revealed that five of the variables exhibit circularly polarised emission, while there are no further circularly polarised sources in the field}. Seven radio sources are associated with known objects -- six stars and one pulsar. While the purpose of the DWF program is to have simultaneous multi-wavelength observations of the sky, the size mismatch between the DECam and ASKAP fields of view means that \fstar\ is the only variable radio source with simultaneous DECam observations. \gaiasource\ is within the DECam field of view, but lies in a chip gap.

\begin{figure*}
    \centering
    \includegraphics{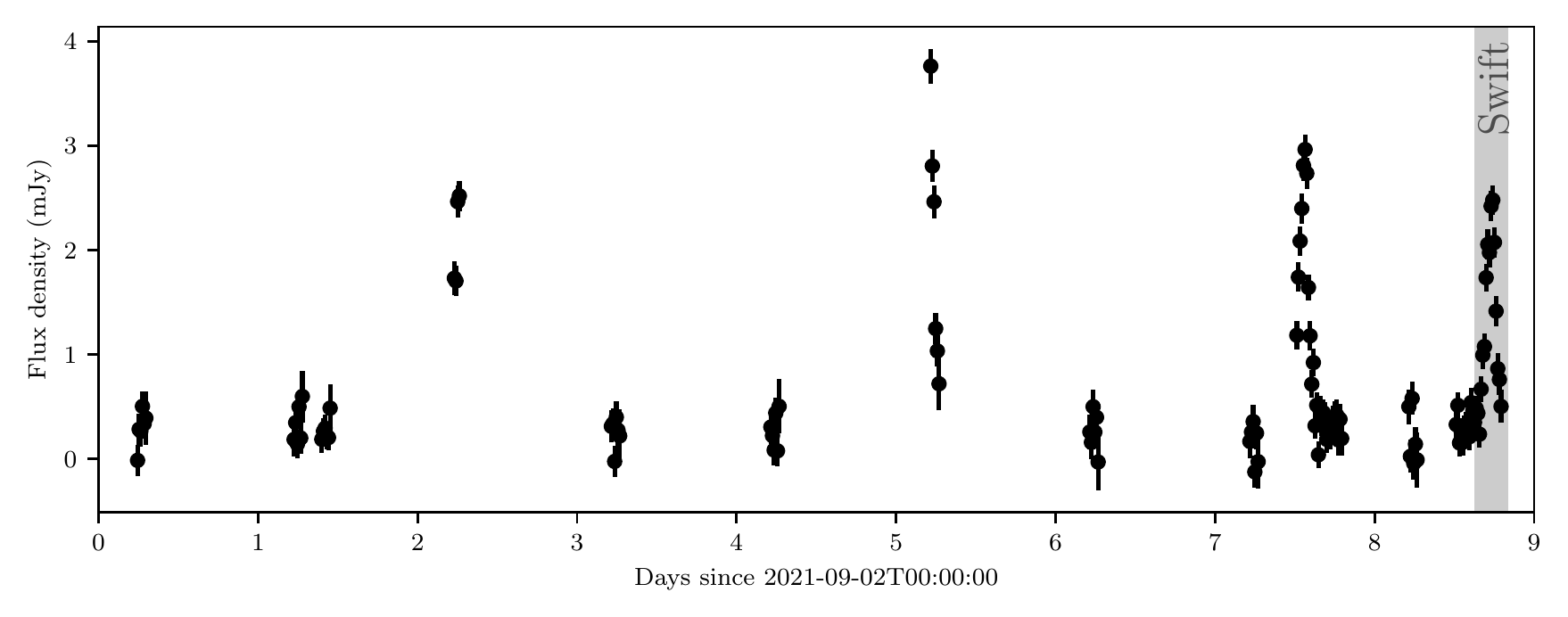}
    \caption{Intra-observation light curve of \fred, using 15-minute samples created as part of the VAST fast-imaging search (Section \ref{subsec:fast-search}). We have excluded the first and last observations which were carried out long before/after the other observations. The shaded region denotes the duration of the simultaneous {\it Swift} observation, in which no X-ray or UV emission was detected.}
    \label{fig:fred_lc}
\end{figure*}

\begin{figure}
    \centering
    \includegraphics{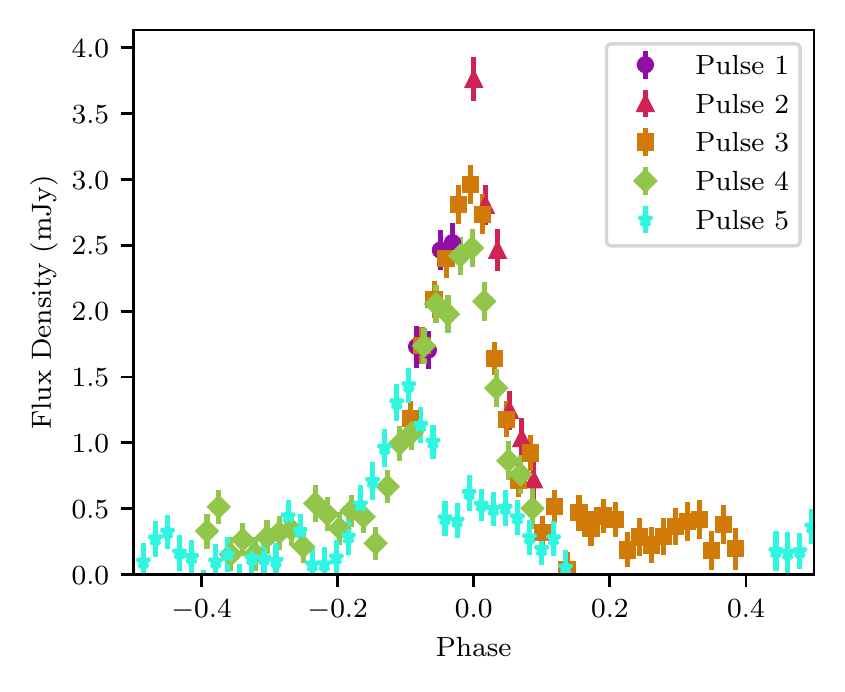}
    \caption{\changes{Intra-observation light curves of \fred\ from the five epochs where we detected a pulse, phased to the 14.1\,hr optical period measured using TESS data.}}
    \label{fig:fred_lsp}
\end{figure}

\subsection{\fred}
\label{subsec:speedy_fred}

\fred\ is an M8.5-dwarf/T6-dwarf binary system \citep{2007A&A...471..655K} with a parallax of $249.7\pm0.1$\,mas and a proper motion of 2.6\,\arcsec per year \citep[][]{2016A&A...595A...1G,2022arXiv220800211G}. It was first noted as a source of interest through the VAST pipeline search using data up to and including the first epoch of the observing run (SB31700) and was later also detected by both the polarisation and fast-imaging searches. The radio light curve (Figure \ref{fig:star_light_curves}) is characterised by a series of marginal detections of $\sim 0.5\,$mJy with four observations showing an increased flux density of either $\sim 3$\,mJy or $\sim 1$\,mJy. Based on the first two detections of $\sim 3$\,mJy, we triggered additional ASKAP observations on 2021-09-09 (SB32018), 2021-09-10 (SB32039) and 2021-09-19 (SB32235) to characterise the observed behaviour.

Figure \ref{fig:fred_lc} shows the intra-observation light curve, from which we conclude that the variability stems from \changes{multiple pulses, each lasting} $\sim 2$\,h. The radio emission is highly ($\sim$90\%) circularly polarised during the pulses, but we do not detect any polarised emission from the source in its quiescent state. We detected a fifth pulse in the final epoch that is not visible in the full light curve (Figure \ref{fig:star_light_curves}) because of the long integration time, which averages it out across five times the pulse duration.

\fred\ is known to be variable at X-ray wavelengths, and $L_{X} = 8\times 10^{27}$\,erg\,s$^{-1}$ flares have previously been detected from it \citep[][]{2010A&A...513A..12R}. We obtained a 3.5 ks exposure with \textit{Swift} (ToO request \#16280) from 2021-09-10 15:13:02 to 2021-09-10 19:53:53 to search for contemporaneous radio--X-ray flares.  We detected 19 photons spatially coincident with \fred\ over the full observation, \changes{but} we were unable to obtain a spectra. The count rate is consistent with the quiescent flux observed by \citet[][]{2010A&A...513A..12R}. No emission was detected with the UV instrument (UVOT). Noting that this observation overlaps with the fourth pulse detected with ASKAP \changes{(see Fig. \ref{fig:fred_lc})}, we conclude that the X-ray and radio activity are not coupled.

\fred\ was observed with TESS during Sector 13 and we measure an optical period of $14.1\pm0.2\,$h after manually extracting the light curve at the proper-motion corrected source location. \changes{We also performed a periodicity search on the intra-observation light curve (see Appendix \ref{app:fred_period}) and measure a radio period of $14.2\pm 0.3$\,h\footnote{This measurement is also in good agreement with preliminary analysis of follow-up observations conducted with the Australia Telescope Compact Array (ATCA) that will be presented in a separate manuscript.}.} 

\changes{Figure \ref{fig:fred_lsp} shows the intra-observation light curves for the five epochs with detected pulses folded to the TESS period, showing that the pulses are all aligned to the same rotational phase, and suggesting that the variability at each wavelength shares a common origin. While \fred\ is known to have a binary companion, the orbital separation is approximately 4\,AU \citep[][]{2007A&A...471..655K}. The corresponding orbital period is therefore on the order of decades and hence we rule out the periodicity originating from the binary orbit. We also rule out the radio variability originating from an interaction between the two stars for similar reasons. The periodic signal in the TESS light curve is quasi-sinusoidal and consistent with the signature of spots on a rotating star. The radio light curve is also consistent with pulses from a rotating star \citep[e.g. comparable to UV Ceti;][]{2019MNRAS.483..614Z,2022ApJ...935...99B}. We cannot conclusively rule out the pulses being produced by the interaction between the star and an exoplanet (or an undetected low-mass stellar companion) as this scenario would produce periodic, circularly polarised radio pulses similar to what we observe \citep[e.g.][]{2007P&SS...55..598Z}. However, this and other similarly exotic scenarios are unlikely given the behaviour of the star at optical wavelengths.}

We conclude that the observed radio variability stems from rotational modulation, making this the slowest rotating radio-pulsing ultra-cool dwarf discovered to-date. It is also one of only a small number of ultra-cool dwarfs detected at sub-gigahertz frequencies \citep{2019MNRAS.483..614Z,2020ApJ...903L..33V}. Further analysis of this source, including the results of our ongoing monitoring campaign, will be presented in a future manuscript.

\subsection{\fstar}
\fstar\ is a known cool dwarf \citep[likely M5V;][]{2019AJ....158..138S} with a parallax of $13.32\pm0.03$\,mas \citep[][]{2021A&A...649A...1G}. We measure an optical period of $9.5\pm1.4$\,days after combining data from TESS sectors 13 and 27. It is the only object in our sample that was covered by simultaneous DECam observations. 

\fstar\ is detected in a single ASKAP observation (SB31751; 2021-09-05) with a flux density of $330\pm 50\,\mu$Jy. We also measure a fractional circular polarisation of $\sim 60$\%. The intra-observation light curve for that epoch shows no significant variability, suggesting that the outburst lasted longer than the observation duration of approximately 200 minutes. It is not detected in the intra-observation imaging of any other epoch.

\begin{figure*}
    \centering
    \includegraphics{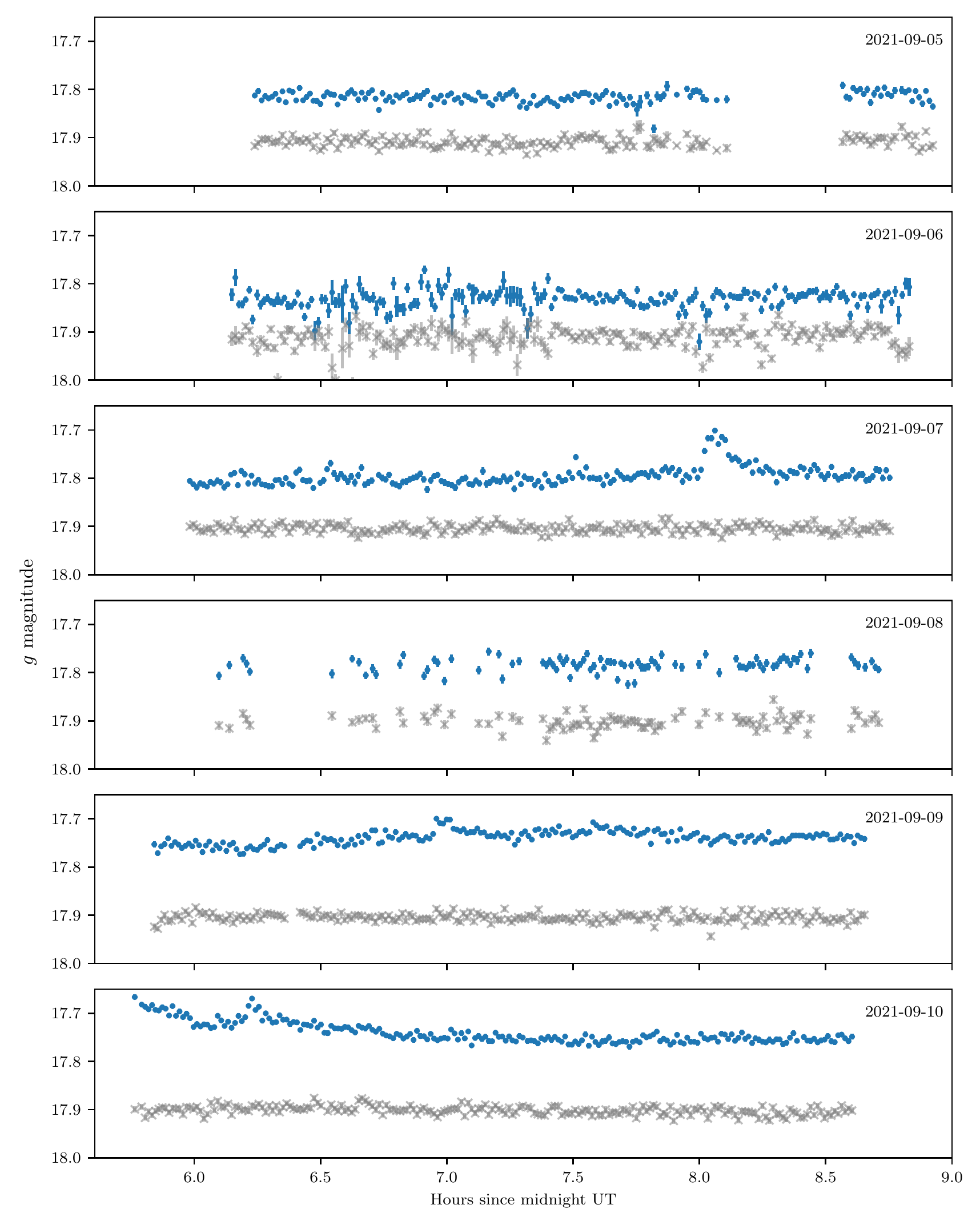}
    \caption{Optical light curves of \fstar\ (blue circles) and an unrelated reference source in the field (grey crosses) across the six nights of the observing run. Observations were carried out with the Dark Energy Camera using the $g$ band filter and there were simultaneous ASKAP observations on all nights. The radio emission from \fstar\ was detected on the first night of DECam observing (2021-09-05). We find strong evidence that \fstar\ is variable at optical wavelengths, but rule out any correlation with the observed radio variability.}
    \label{fig:fstar_decam_lc}
\end{figure*}

Figure \ref{fig:fstar_decam_lc} shows the DECam light curve across the six nights of the run, showing that the source exhibits minimal variability on the observed timescales. We see a small ($\Delta m \approx 0.1$) flare on 2021-09-07 at 08:00\,UTC, and marginal evidence for smaller flares on 2021-09-09 and 2021-09-10, but no unusual behaviour during the period surrounding the radio detection. Further radio observations are required to determine whether the radio emission is due to rotational modulation, as our observations sample less than two rotations of the star.

\subsection{\cfoct}
\cfoct\ is a likely RS CVn star \citep[e.g.][]{1989JApA...10..139P,2012MNRAS.420.2539B} with a parallax of $4.45\pm0.06$\,mas \citep[][]{2021A&A...649A...1G}. \citet{2012MNRAS.420.2539B} found a rotational period of 19.90--20.45 days, with the large range caused by differential rotation on the stellar surface. The star is known to be active at radio wavelengths. \citet{1987PASA....7...55S,1987MNRAS.229..659S} reported the detection of circularly polarised bursts at 8.4\,GHz reaching tens of mJy. It has also been detected with a flux density of 2\,mJy at 843\,MHz \citep{1987PASA....7...42V}, broadly consistent with our measurements. The source exhibits significant radio inter-observation variability ($\eta=78.5$, $V=0.42$), but does not appear variable on 15-minute timescales except for a slight rise across the final observation. We also do not detect any circularly polarised emission, despite it being it being common in radio-loud RS CVn systems \citep[][]{2003A&A...403..613G}.

The radio light curve morphology is of particular interest, with no clear trend in the first four measurements followed by a smooth, quasi-parabolic, decline and rise (which we henceforth refer to as ``the dip''). We are unable to conclusively determine the origin of this variability with existing observations. However, there are a number of plausible scenarios which could explain the observed behaviour.

The dip is not consistent with a simple sinusoid at the optical rotation period (and is instead better fit by a period of $\sim 12$\,d), but we cannot rule out the variability being caused by some form of rotational modulation. For example, if the star has a large number of magnetically active regions distributed unevenly across the surface, the observed light curve may be explained by a particularly under-active region rotating in and out of view \citep[e.g.][]{Pakhomov2015}. 

While \cfoct\ has a binary companion, the system is known to be non-eclipsing \citep[][]{1989JApA...10..139P}, and hence we rule out an eclipse as the origin of the dip. The dip may originate from some form of orbital interaction (e.g. direct orbital modulation, or simply decreased emission near apoapsis), although we note that RS CVn are typically tidally locked \citep[][]{2015AstL...41..677P} and hence a simple explanation along these lines is disfavoured for similar reasons as above.

Ultimately, further interpretation of this source is limited by our relatively short observing campaign which does not cover a single rotational/orbital period. A dedicated observing campaign spanning at least a month will shed light on the cause of this intriguing variability, while Very Long Baseline Interferometry would be useful in determining whether the radio emission originates from one star in the system or the binary separation region.

\begin{figure}
    \centering
    \includegraphics{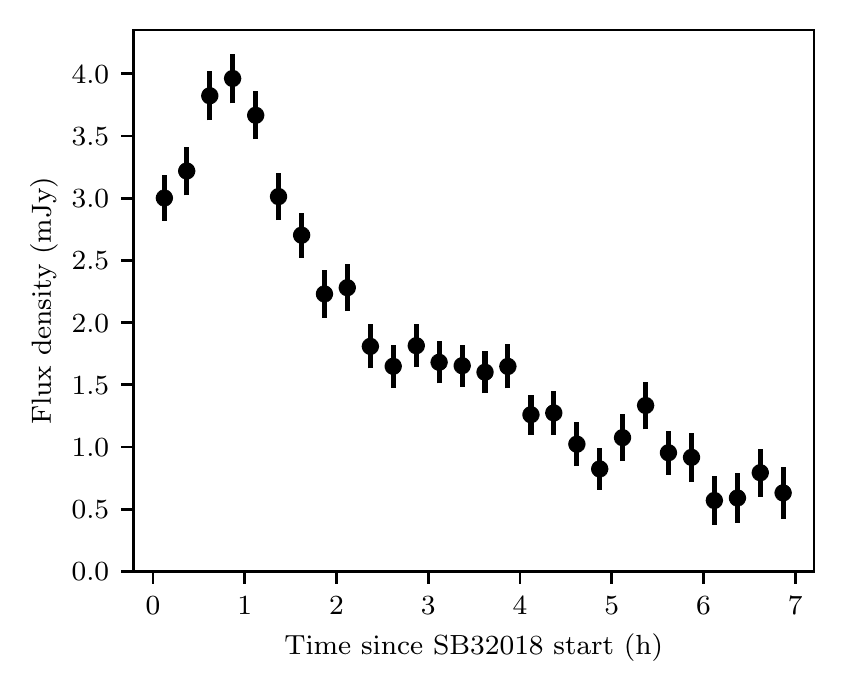}
    \caption{Intra-observation light curve of \pulsar\ from SB32018 using 15-minute samples generated as part of the VAST fast-imaging search (see Section \ref{subsec:fast-search}). The observed variability is likely caused by diffractive scintillation.}
    \label{fig:psrj1933_fast_lc}
\end{figure}

\subsection{\pulsar}
\pulsar\ is a millisecond binary pulsar with a spin period of $\sim 3.54$\,ms and a dispersion measure of 11.52\,pc\,cm$^{-3}$ \citep{2007ApJ...656..408J,2017MNRAS.471.4579G}. It was noted as a source of interest in the VAST pipeline search and we measure significant variability in the full light curve, with $\eta=83.4$ and $V=0.64$. \pulsar\ has been observed to exhibit variability due to diffractive scintillation at 1.3\,GHz with a timescale of $\sim 2$\,h \citep{2017MNRAS.471.4579G}, longer than most of our observations of this field. Figure \ref{fig:psrj1933_fast_lc} shows the intra-observation light curve of this source from SB32018 where we measure flux densities spanning 0.5--4\,mJy across 7\,h of observation with a characteristic timescale broadly in agreement with that reported by \citet{2017MNRAS.471.4579G}. \pulsar\ exhibits similar variability in the two other long observations (SB32039 and SB32235), albeit with a lower average flux density. We therefore conclude that the observed variability is solely due to diffractive scintillation.

\begin{figure*}
    \centering
    \includegraphics{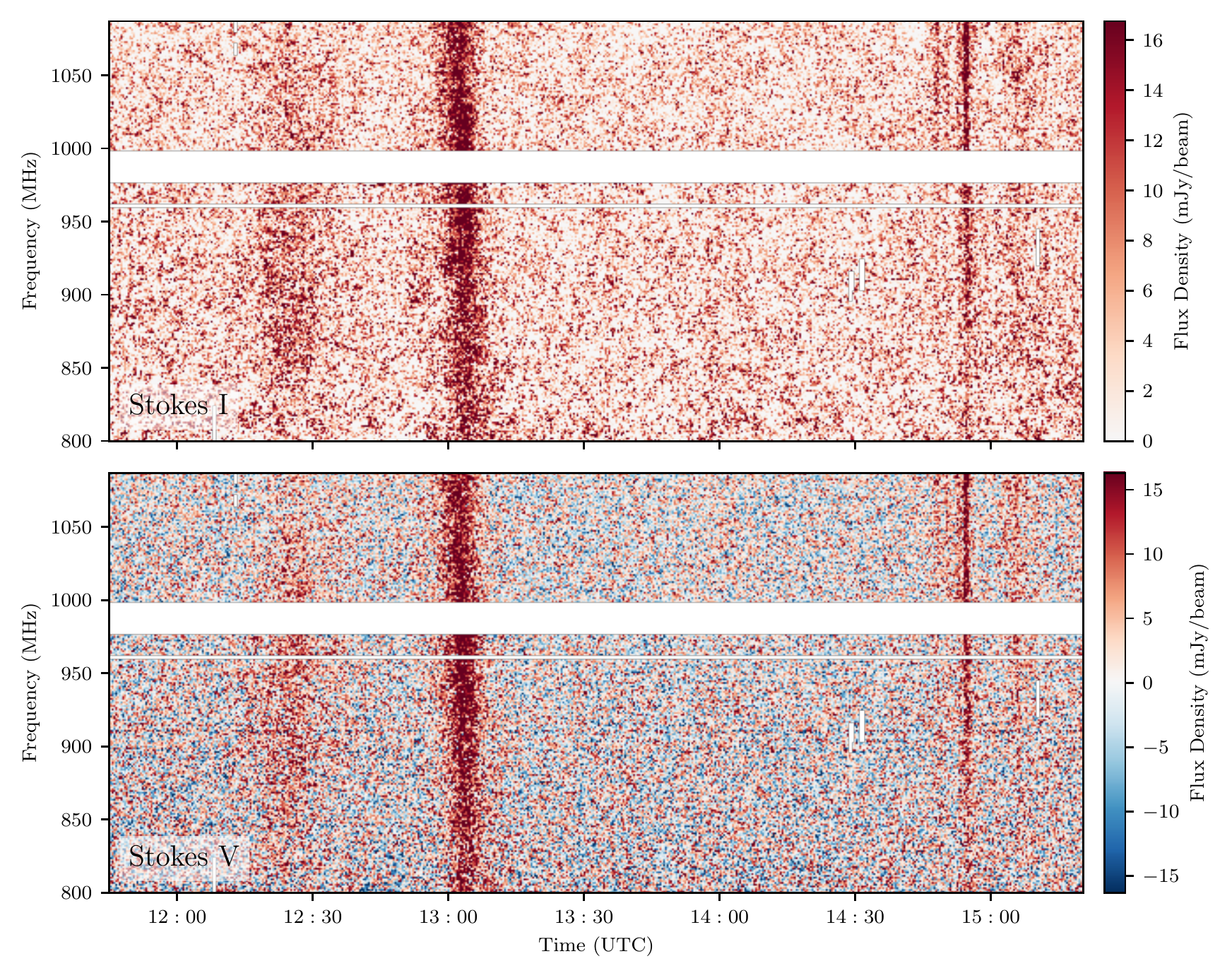}
    \caption{Dynamic spectra of \cdstar\ showing the detection of four circularly polarised bursts on 2021-09-19. Top: Stokes I. Bottom: Stokes V.}
    \label{fig:cd64_ds}
\end{figure*}

\subsection{\cdstar}
\label{subsec:cdstar}
\cdstar\ is a K5Ve star \citep[][]{2006A&A...460..695T} with a parallax of $35.2\pm0.2$\,mas \citep{2021A&A...649A...1G} and a rotation period of $8.5\pm 0.1$\,h \citep{2011A&A...533A..30G}.

The star was first noted as a source of interest in the VAST pipeline search where we measure $\eta=169$ and $V=0.4$. The first 13 epochs of the inter-observation light curve \changes{(Fig. \ref{fig:star_light_curves})} show a small degree of variability, consistent with slow variations in the flux of quiescent emission. It exhibits no clear variability in the intra-observation light curves except for the final epoch (SB32235) where we detect a highly circularly polarised ($\sim 100\%$) quadruple outburst spanning approximately 3\,h. This part of the rotational phase is covered by multiple other epochs and we find no evidence for comparable behaviour in any of them.

To study the structure of these bursts in more detail we created dynamic spectra for all Stokes parameters. We phase rotated the model-subtracted visibilities (generated during the fast-imaging search; see Section~\ref{subsec:fast-search}) from the beam center to the coordinates of \cdstar. To minimize the influence of poorly-modelled diffuse emission on the dynamic spectra, we vector-averaged visibilities for each instrumental polarization only across baselines longer than 200\,m. We combined the complex visibilities for each instrumental polarization to produce dynamic spectra for full Stokes parameters. 

Figure \ref{fig:cd64_ds} shows the resulting Stokes I and V dynamic spectra (we do not show Stokes Q or U as the source is not detected). 
We find that the bursts show frequency structure, with the first burst being band-limited between 850--1050\,MHz and the final two having a low-frequency cutoff of $\sim 900\,$MHz. The second burst exhibits a frequency drift of approximately $-1.1\,$MHz/s, similar to that observed in pulses from UV Ceti, TVLM 513--46, and 2M\,0746+20 \citep[][]{2015ApJ...802..106L,2019MNRAS.488..559Z}. However, in those instances the pulses and drift were attributed to rotationally modulated auroral emission produced by the electron cyclotron maser instability \citep[ECMI;][]{2006A&ARv..13..229T}, with pulses detected at the same rotational phase over multiple stellar rotations. In the case of \cdstar\ we have comprehensive coverage of the full stellar rotation period and only detect the four pulses in SB32235, with a 9 day gap separating the previous observation at the same rotational phase in SB32039. Hence, if this variability is rotationally modulated the local plasma conditions must have only become favourable for ECMI to operate in the previous 9 days. Alternatively, the four bursts may be driven by stochastic stellar magnetic activity \citep[e.g.][]{Villadsen2019}. \changes{Further monitoring will distinguish between these competing scenarios.}

\begin{figure}
    \centering
    \includegraphics{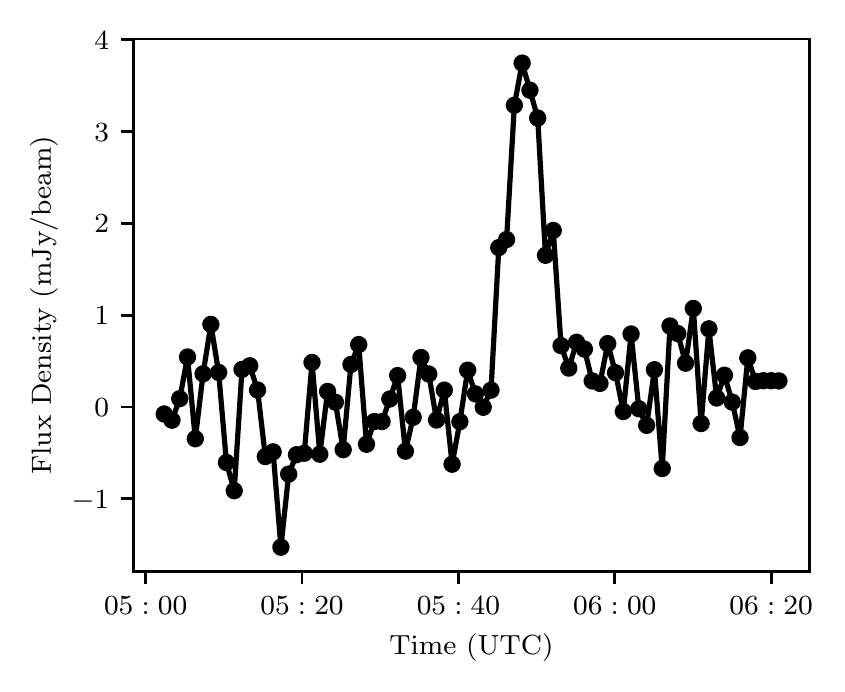}
    \caption{Intra-observation light curve of \gaiasource\ from SB31750 using 1-minute samples, showing that the radio emission originates from a flare lasting approximately 10\,minutes.}
    \label{fig:gaia_source_1minute_lc}
\end{figure}

\subsection{\gaiasource}
This source has only previously been catalogued by Gaia, likely due to its proximity to \tyc, a 10th magnitude star only 4\,\arcsec away. It has a parallax of $2.27\pm0.05$\,mas, and is classified as a likely cataclysmic variable with 84\% confidence in Gaia DR3.

\gaiasource\ is only detected in a single ASKAP observation (SB31750; 2021-09-05) where we measure a peak flux density of $510\pm 60\,\mu$Jy with a fractional circular polarisation of 98\%. The intra-observation light curve for that epoch shows a single integration detection of $1.6\pm0.1$\,mJy. We created a higher time resolution light curve following the same procedure used to generate the dynamic spectra of \cdstar\ (Section \ref{subsec:cdstar}). To achieve a good signal-to-noise we then averaged across all sub-bands and every six integrations, resulting in a temporal resolution of 1\,minute.

The resulting light curve (Fig. \ref{fig:gaia_source_1minute_lc}) shows a single burst lasting approximately 10 minutes. Accounting for the possibility of beamed emission, we infer a peak radio luminosity of $\sim 1\times 10^{18}$\,erg\,s$^{-1}$\,Hz$^{-1}$, consistent with other detections of stars with ASKAP (Pritchard et al., in prep.).

\subsection{\askapj}
\askapj\ is a previously uncatalogued radio source that exhibits a gradual rise across the course of the observing period \changes{(Fig. \ref{fig:star_light_curves})}, with $\eta=3.5$, $V=0.17$. The source shows negligible intra-observation variability except for SB31702 where it increases from a typical flux density of $\sim 0.5$\,mJy \changes{(ranging from 0.3--0.9\,mJy)} to $2.0\pm 0.3$\,mJy in the final \changes{30 minutes} of the observation.

The source is located close to the known variable star \ftoct, but based on the light curve morphology and the 6.1\arcsec offset between the radio and proper-motion corrected optical positions, it is likely unrelated. However, the proximity to such a bright optical/infra-red source means that we are unable to place any constraints on this source using archival multi-wavelength data. 

We can use the NE2001 model of galactic electron density \citep[][]{2002astro.ph..7156C} and the equations of \citet[][]{1998MNRAS.294..307W} to determine the expected properties of scintillation-induced variability for extragalactic compact sources. We estimate a characteristic timescale of $\sim 17$\,days and a characteristic amplitude of $\sim 20\%$ along the line of sight to \askapj\ at 943\,MHz, in good agreement with the observed radio light curve. \changes{We note that while there is clear correlation in flux density between consecutive observations, this is not unexpected if the variability is due to refractive scintillation, since the observing cadence is much less than the scintillation timescale.} We therefore propose that this source is likely a scintillating extragalactic source rather than an intrinsically variable source.

\begin{figure}
    \centering
    \includegraphics{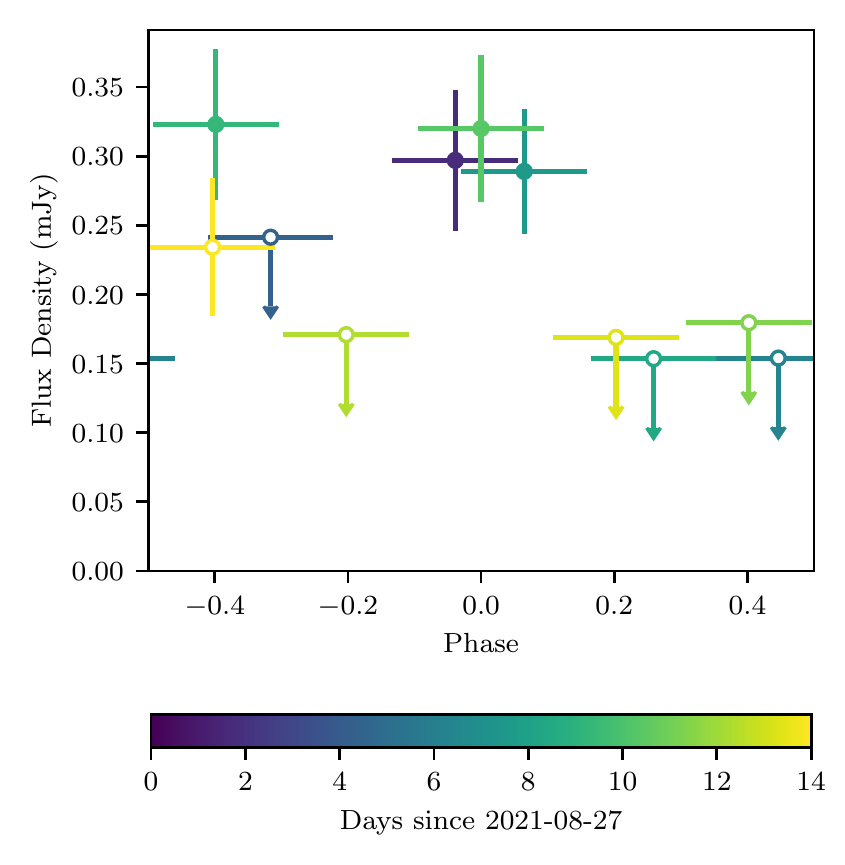}
    \caption{Radio light curve of \ucac053\ folded to the optical period of 17.2\,h. The horizontal errorbars denote the duration of the observation, and non-detections are shown as $3\sigma$ upper limits. The marginal detection in SB32036 is denoted by an open marker.}
    \label{fig:ucac_053_folded}
\end{figure}

\subsection{\ucac053}
\label{subsec:ucac053}
\ucac053\ is a known cool dwarf \citep[likely M4V;][]{2019AJ....158..138S} with a parallax of $14.44 \pm 0.02$\,mas \citep[][]{2021A&A...649A...1G}. We measure an optical period of $17.16 \pm 0.20$ days after combining data from TESS sectors 13 and 27.

\ucac\ was initially found by the VAST pipeline search and has variability metrics of $\eta=3.77$ and $V=0.55$. The light curve \changes{(Fig. \ref{fig:star_light_curves})} is characterised by four $\sim 0.3$\,mJy detections, and a fifth marginal detection with similar flux density. Each detection is consistent with a polarisation fraction of $\sim 70\%$, although the detection of Stokes V emission in SB31377 and SB31751 is marginal. The source is not detected in any intra-observation image due to insufficient sensitivity, ruling out this variability originating from short-timescale flares. Figure \ref{fig:ucac_053_folded} shows the inter-observation light curve folded to the optical period, and we note that the detections occur in two groups separated by $140\deg$ of rotational phase.

\changes{Each of the eleven observations of this source covers a rotational phase that overlaps with at least two other observations (i.e. there are 11 unique pairs of overlapping observations). There are four intervals of rotational phase covered by three observations. To assess the significance of the grouping of the detections we assume that the detections are unrelated to the rotation of the star, and hence would be distributed randomly in rotational phase. There are ${11 \choose 5}=462$ ways to distribute five detections between our eleven observations. Of these, 104 (23\%) have at least three detections that overlap in phase space, 65 (14\%) have all five detections overlapping with at least one other detection, and 52 (11\%) arranged into exactly one group of two overlapping detections and one group of three overlapping detections. We therefore conclude that the probability of a similar scenario occurring by chance is greater than $11\%$. Hence, the observed clustering is not statistically significant, but does warrant further investigation.}

\changes{If the grouping of the detections is real, o}ne possible explanation for the observed \changes{behaviour} is that the detections originate from two independent active regions on the star -- one which has persisted throughout the observation period, and another that arose partway some time after 2021-09-01. On the other hand, the comparable polarisation fraction of all five detections perhaps suggests a single origin, independent of the star's rotation. However, we were unable to find a period that aligns all five detections and does not also align them with constraining non-detections. Given the relatively low significance of each detection, the observed variability could also be explained by low level quiescent variability \changes{either associated with the rotation of the star, or intrinsic to it}. Ultimately, we are unable to definitively determine the origin of the variability from these observations alone.

\section{Discussion and Summary}
We have carried out a search for transient, variable and circularly polarised radio sources using daily observations of two fields with \changes{ASKAP}. We used a three-pronged approach consisting of a standard variability search of the default images produced by the telescope's data reduction pipeline (see Section \ref{subsec:pipeline-search}), an examination of all sources detected in the Stokes V images (see Section \ref{subsec:pol-search}), and a ``fast-imaging'' approach which involved splitting each observation into 15-minute sub-integrations (see Section \ref{subsec:fast-search}).

We found eight sources of interest in total -- all eight in the standard variability search and five in each of the polarisation and fast-imaging searches \changes{as shown in Table \ref{tab:variables}}. \changes{Of these, six are intrinsically variable emission from stellar systems, one is a pulsar exhibiting diffractive scintillation and one is likely an extragalactic source exhibiting refractive scintillation. Three of the stellar systems are known M-dwarfs, one is a K-dwarf, one is an RS CVn while the spectral type of the sixth is unknown. Of particular interest is the detection of circularly polarised bursts from three of the stars (\fred, \gaiasource and \cdstar) lasting minutes--hours. We have determined that the variable emission from \fred\ originates from the rotation of the star, but the mechanism by which the other five stellar systems are variable remains unclear.}

Our approach demonstrates the utility of using a variety of techniques when searching for and classifying sources of interest in an untargeted widefield search. For example, determining that the emission from \fred\ was periodic required the results of the fast-imaging search even though it was initially detected by the standard variability search. Similarly, if our observing cadence had been different \ucac\ may not have been found in the standard variability search (e.g. all observations made detections with similar flux densities) although it would have been detected by the polarisation search.

These observations probe a new part of the gigahertz-frequency transient parameter space -- previous widefield searches have generally had an observing cadence of months--days, while searches with a comparable cadence have lacked areal coverage. Our combination of widefield coverage and high cadence makes this survey sensitive to variability on timescales of days while also producing a reasonably large event rate. Furthermore, the high cadence was also vital for inferring source properties from the observed light curves. For example, the daily cadence enabled us to measure the period of the pulses from \fred\ to a precision of less than one hour, and in turn allowed us to conclusively link the observed variability to the rotation of the star.

\changes{While seven of the eight sources have been previously catalogued by surveys at other wavelengths, this is the first time variable radio emission has been detected from most of them. More importantly, only \fred\ (producing large X-ray flares) and \cfoct (producing radio flares) were known to be particularly interesting prior to this search -- the other five known objects show no remarkable features that would motivate further dedicated study of them. The widefield nature of these observations means that our search is not biased by selection effects such as those encountered by targeted searches for radio variability from known objects. This work, and other widefield searches, therefore present a unique opportunity to study topics such as the link between stellar radio variability and activity at other wavelengths.}

ASKAP is uniquely placed to carry out searches for bright short-timescale transients like those presented in this work as it is the only widefield gigahertz-frequency interferometer currently operating. Similar searches can be carried out with MeerKAT, but will be hindered by the comparably small field of view. The relatively large number of variable sources we have detected motivates a dedicated observing campaign independent of the planned VAST survey (which will not achieve comparable sensitivity or cadence). However, any such survey should target the Galactic plane to increase the prospects of detecting emission from GCRTs, magnetars and pulsars. We note that the planned DSA-2000 \citep[][]{2019BAAS...51g.255H} will be well-placed to carry out similar searches.

\section*{Acknowledgements}
Parts of this research were conducted by the Australian Research Council Centre of Excellence for Gravitational Wave Discovery (OzGrav), project number CE170100004. This work was partially supported by International Partnership Program of Chinese Academy of Sciences (Grant No.113111KYSB20190020).

The Australian SKA Pathfinder is part of the Australia Telescope National Facility (grid.421683.a) which is managed by CSIRO. Operation of ASKAP is funded by the Australian Government with support from the National Collaborative Research Infrastructure Strategy. ASKAP uses the resources of the Pawsey Supercomputing Centre. Establishment of ASKAP, the Murchison Radio-astronomy Observatory and the Pawsey Supercomputing Centre are initiatives of the Australian Government, with support from the Government of Western Australia and the Science and Industry Endowment Fund. We acknowledge the Wajarri Yamatji people as the traditional owners of the Observatory site.

This work has made use of data from the European Space Agency (ESA) mission
{\it Gaia} (\url{https://www.cosmos.esa.int/gaia}), processed by the {\it Gaia}
Data Processing and Analysis Consortium (DPAC,
\url{https://www.cosmos.esa.int/web/gaia/dpac/consortium}). Funding for the DPAC
has been provided by national institutions, in particular the institutions
participating in the {\it Gaia} Multilateral Agreement.

This paper includes data collected by the TESS mission, which are publicly available from the Mikulski Archive for Space Telescopes (MAST). Funding for the TESS mission is provided by NASA's Science Mission directorate.

This project used data obtained with the Dark Energy Camera (DECam), which was constructed by the Dark Energy Survey (DES) collaboration. Funding for the DES Projects has been provided by the US Department of Energy, the US National Science Foundation, the Ministry of Science and Education of Spain, the Science and Technology Facilities Council of the United Kingdom, the Higher Education Funding Council for England, the National Center for Supercomputing Applications at the University of Illinois at Urbana-Champaign, the Kavli Institute for Cosmological Physics at the University of Chicago, Center for Cosmology and Astro-Particle Physics at the Ohio State University, the Mitchell Institute for Fundamental Physics and Astronomy at Texas A\&M University, Financiadora de Estudos e Projetos, Funda\c c\~ao Carlos Chagas Filho de Amparo \`a Pesquisa do Estado do Rio de Janeiro, Conselho Nacional de Desenvolvimento Cient\'ifico e Tecnol\'ogico and the Minist\'erio da Ci\^encia, Tecnologia e Inova\c c\~ao, the Deutsche Forschungsgemeinschaft and the Collaborating Institutions in the Dark Energy Survey.

The Collaborating Institutions are Argonne National Laboratory, the University of California at Santa Cruz, the University of Cambridge, Centro de Investigaciones En\'ergeticas, Medioambientales y Tecnol\'ogicas-Madrid, the University of Chicago, University College London, the DES-Brazil Consortium, the University of Edinburgh, the Eidgen\"ossische Technische Hochschule (ETH) Z\"urich, Fermi National Accelerator Laboratory, the University of Illinois at Urbana-Champaign, the Institut de Ci\`encies de l`Espai (IEEC/CSIC), the Institut de F\'isica d`Altes Energies, Lawrence Berkeley National Laboratory, the Ludwig-Maximilians Universit\"at M\"unchen and the associated Excellence Cluster Universe, the University of Michigan, NSF's NOIRLab, the University of Nottingham, the Ohio State University, the OzDES Membership Consortium, the University of Pennsylvania, the University of Portsmouth, SLAC National Accelerator Laboratory, Stanford University, the University of Sussex, and Texas A\&M University.

Based on observations at Cerro Tololo Inter-American Observatory, NSF's NOIRLab (NOIRLab Prop. ID 2020B-0253; PI: J. Cooke), which is managed by the Association of Universities for Research in Astronomy (AURA) under a cooperative agreement with the National Science Foundation.

This research has made use of the SIMBAD database, operated at CDS, Strasbourg, France. This research has made use of NASA's Astrophysics Data System. 

This research made use of Lightkurve, a Python package for Kepler and TESS data analysis \citep[][]{2018ascl.soft12013L}.

This research made use of matplotlib, a Python library for publication quality graphics \citep{Hunter:2007}, SciPy \citep{Virtanen_2020} and Astropy, a community-developed core Python package for Astronomy \citep{2013A&A...558A..33A,2018AJ....156..123A}. 

\section*{Data Availability}
The ASKAP data used in this work can be accessed through the CSIRO ASKAP Science Data Archive (CASDA\footnote{\url{https://data.csiro.au/dap/public/casda/casdaSearch.zul}}) using the SBIDs listed in Table \ref{tab:obs_descrip}. The DECam and HXMT data will be shared on reasonable request to the corresponding author.

\appendix

\section{Measuring the period of \fred}
\label{app:fred_period}
\begin{figure}
    \centering
    \includegraphics{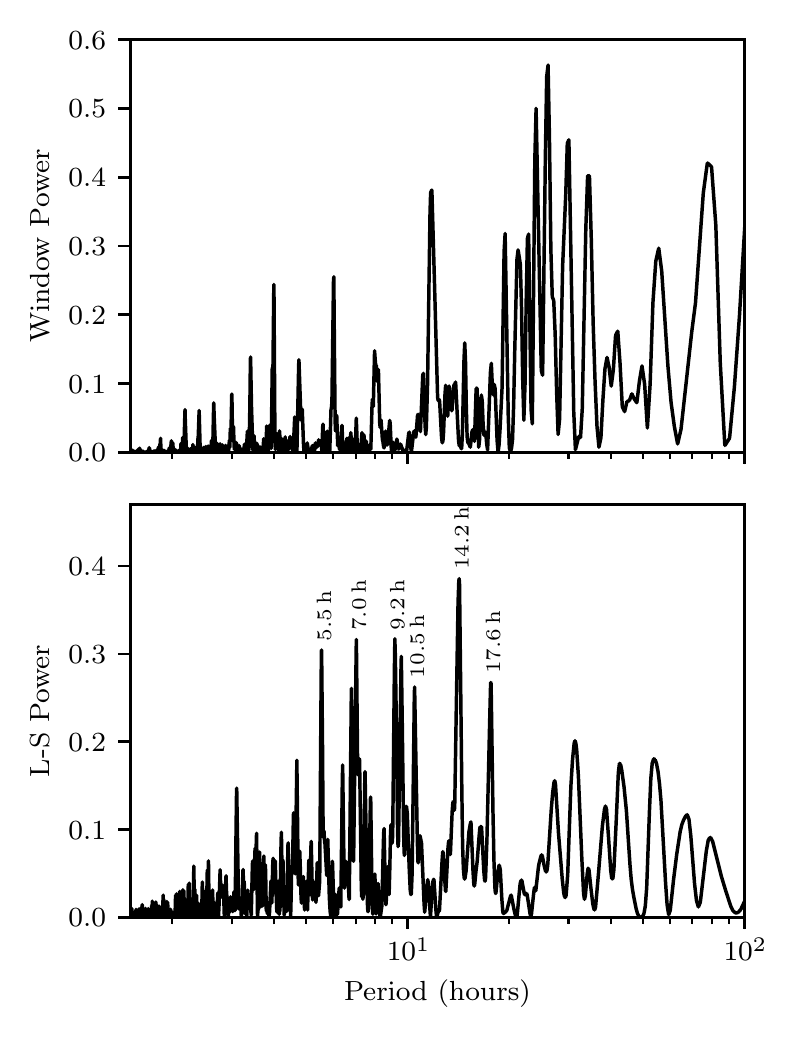}
    \caption{Top: Window function of our observations of \fred. Bottom: Lomb-Scargle periodogram of the intra-observation light curve of \fred with the periods corresponding to prominent peaks labelled.}
    \label{fig:fred_ls_window}
\end{figure}

To search for periodicity in the emission observed from \fred\ we followed the recommendations outlined by \citet[][]{2018ApJS..236...16V}. We used \textsc{astropy.timeseries.LombScargle} to calculate the Lomb-Scargle periodogram of the intra-observation light curve as well as the associated window transform (Figure \ref{fig:fred_ls_window}). We set \textsc{samples\_per\_peak=10}, but all other options were left default.

The window transform shows a large amount of structure, with significant peaks at 24, 26, 30 and 34 hours. We also note the presence of a peak at 22.7 hours, corresponding to the gap between the end of one day's observations and the beginning of the next. All peaks in the window function below 15 hours occur at periods of $24/n$ for integer $n$ and are aliases of the 24 hours spike. This level of structure is to be expected given our non-uniform observing windows.

The periodogram power is maximum at a period of $14.2\pm0.3$\,h, but we also find five other significant peaks at periods of $17.6\pm 0.3$, $10.5\pm 0.1$, $9.1\pm 0.1$, $7.0\pm0.1$ and $5.55\pm 0.04$ hours. The presence of multiple peaks is qualitatively consistent with expectations given the highly structuted window transform. We use the Baluev method \citep[][; as implemented in \textsc{astropy}]{2008MNRAS.385.1279B} to calculate the False Alarm Probability (FAP) of each periodogram peak. We find that the 14.2\,h peak has a FAP of $6.5\times10^{-14}$, while the five other peaks have FAPs ranging from $10^{-10}$ to $10^{-8}$. The Baluev method is known to be inaccurate for periodograms with structured window functions, but generally overestimates the FAP and hence, can be treated as an upper limit. We have also attempted to calculate the FAP with a bootstrap method, but found no peaks exceeding the power of those in the periodogram in $10^{5}$ samples, consistent with the upper limit obtained from the Baluev method.

\begin{figure}
    \centering
    \includegraphics{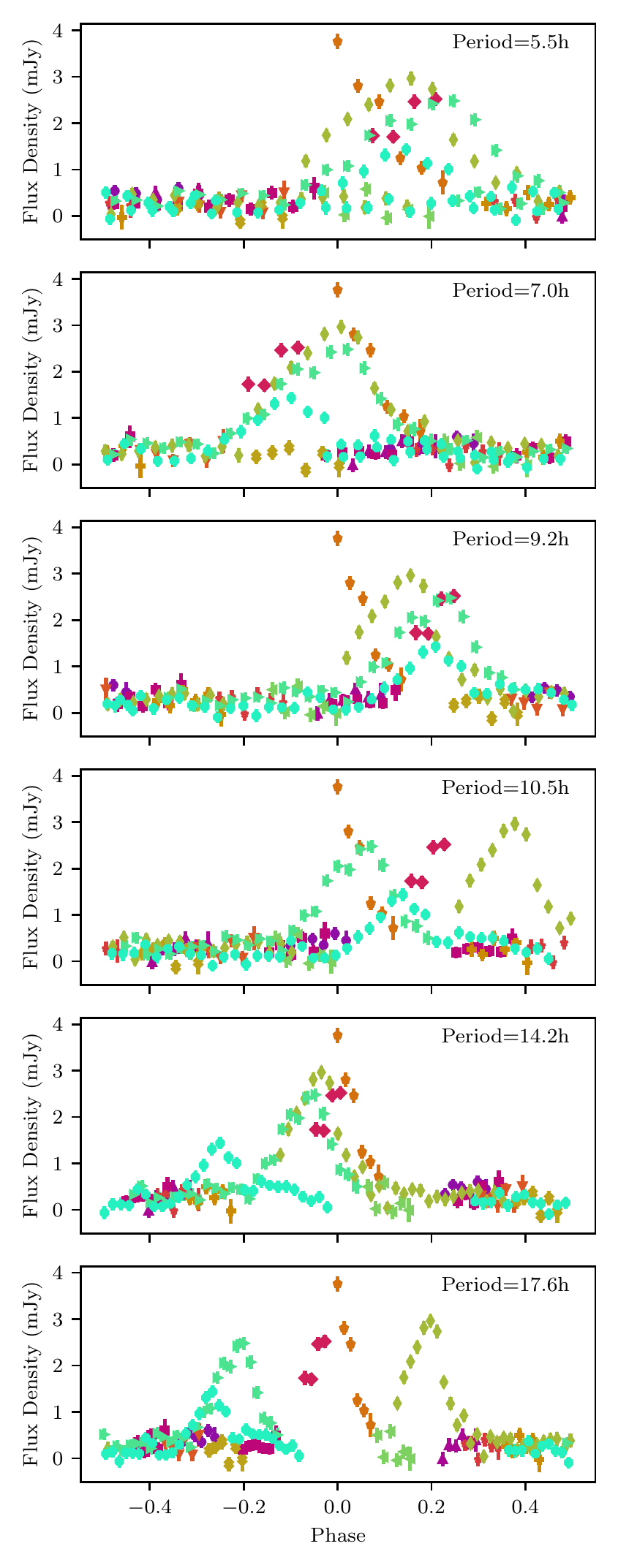}
    \caption{The complete intra-observation light curve of \fred\ folded to the six candidate periods. Only the 14.2\,h light curve shows good alignment between peaks with no constraining non-detections at the pulse phase.}
    \label{fig:speedy_fred_possible_periods}
\end{figure}

There is no generalised method with which to compute the relative likelihood of multiple peaks in the periodogram originating from a real, periodic signal \citep[][]{2018ApJS..236...16V}. Instead, we manually inspect the intra-observation light curves folded to the six candidate periods (Figure \ref{fig:speedy_fred_possible_periods}). Only the light curves folded to 7.0 and 14.2\,h show good alignment between the five detected pulses, but the 7.0\,h light curve also has multiple constraining non-detections at the pulse phase.

We therefore conclude that the true period is $14.2\pm 0.3$\,h with an associated false alarm probability of less than $6.5\times 10^{-14}$. We also note that while the best-fit period measured from the periodogram is $14.2$\,h, there is qualitatively better agreement between detected pulses after folding to the known optical period of $14.1$\,h as shown in Figure \ref{fig:fred_lsp}.

\bsp
\label{lastpage}
\end{document}